\documentclass[cits]{PoS}
\usepackage{epsfig}

\newcommand{\csw}{c_{\rm SW}}
\newcommand{\be}{\begin{equation}}
\newcommand{\ee}{\end{equation}}
\newcommand{\rmO}{{\rm O}}

\newcommand{\fig}[1]{Fig.~\ref{#1}}
\newcommand{\eq}[1]{Eq.~(\ref{#1})}

\title{Non-perturbative improvement of nHYP smeared Wilson fermions}

\ShortTitle{Non-perturbative improvement of nHYP smeared Wilson fermions}
\author{\speaker{Roland Hoffmann}, Anna Hasenfratz\\
        Department of Physics, University of Colorado, Boulder, CO-80309-390, USA\\
        E-mail: \email{hoffmann@pizero.colorado.edu}}
\author{Stefan Schaefer\\
        NIC, DESY, Platanenallee 6, D-15738 Zeuthen, Germany}
        
\abstract{Using Schr\"odinger functional techniques, we determine the coefficient
of the clover term necessary for non-perturbative $\rm{O}(a)$ improvement of hypercubic
smeared Wilson fermions on a quenched plaquette action background.
Unlike for unsmeared Wilson fermions, the resulting clover coefficients are close
to the tree--level value even at coarse lattice spacings,
indicating the absence of large cutoff effects.
A number of exploratory tests are also performed with the improved action.}

\FullConference{The XXV International Symposium on Lattice Field Theory\\
		 July 30 - August 4 2007\\
		 Regensburg, Germany}
		 
\begin{document}

\section{Motivation}
We present results from a non-perturbative determination of the clover coefficient
$\csw$ for Wilson fermions that couple to hypercubic smeared gauge links
\cite{Hasenfratz:2001hp,Hasenfratz:2007rf}. This is done on a quenched gauge background
and while the actual results will have little practical relevance, they provide
an interesting comparison to the unsmeared case. In this way we want to quantify
how much smearing can help with Wilson--type fermions
at coarse lattice spacings, especially concerning their chiral properties.

We use the same gauge background (plaquette gauge action) and technique as
the ALPHA collaboration in their determination of $\csw$ for thin links
\cite{Luscher:1996ug}. The quantitative comparison of smeared and unsmeared
case includes the value of the clover coefficient itself, remnant cutoff effects,
the finite renormalization constant of the vector current
and the range of accessible quark masses at a given lattice spacing.

\section{Setup and strategy}

Simulations of the quenched QCD Schr\"odinger functional (SF) are performed
using the plaquette gauge action and two quenched flavors of improved Wilson fermions.
For unexplained notation regarding the hypercubic (nHYP) smearing or the Schr\"odinger
functional we refer the reader to Refs.~\cite{Hasenfratz:2001hp,Hasenfratz:2007rf,
Luscher:1996ug,Jansen:1995ck,Luscher:1996sc}.
The SF setup imposes Dirichlet boundary conditions in the temporal directions
and thus provides another IR cutoff in addition to the quark mass.
Therefore, in principle, simulations
at the critical point are possible. Moreover, the Schr\"odinger functional is
a convenient formalism to formulate and implement non-perturbative improvement
and renormalization conditions.

The nHYP smearing under consideration here was recently used in dynamical Wilson
clover simulations \cite{Hasenfratz:2007rf,Hasenfratz:2007dc} as well as in the construction
of the kernel for both quenched and dynamical overlap studies \cite{DeGrand:2007tm,
Hasenfratz:2007iv,AnnaProceed}.

When used with Schr\"odinger functional boundary conditions, the HYP construction
\cite{Hasenfratz:2001hp,Hasenfratz:2007rf} has to be modified in the vicinity of
the temporal boundaries. We adopt the prescription that the Dirichlet conditions
are preserved under the blocking such that the original nHYP construction
(labeling lattice sites by $n$)
\begin{eqnarray}
V_{n,\mu} & = & \textrm{Proj}_{U(3)}[(1-\alpha_{1})U_{n,\mu}+\frac{\alpha_{1}}{6}\sum_{\pm\nu\neq\mu}\widetilde{V}_{n,\nu;\mu}\widetilde{V}_{n+\hat{\nu},\mu;\nu}\widetilde{V}_{n+\hat{\mu},\nu;\mu}^{\dagger}]\,,\label{eq:HYP-def1}\\
\widetilde{V}_{n,\mu;\nu} & = & \textrm{Proj}_{U(3)}[(1-\alpha_{2})U_{n,\mu}+\frac{\alpha_{2}}{4}\sum_{\pm\rho\neq\nu,\mu}\overline{V}_{n,\rho;\nu\,\mu}\overline{V}_{n+\hat{\rho},\mu;\rho\,\nu}\overline{V}_{n+\hat{\mu},\rho;\nu\,\mu}^{\dagger}]\,,\label{eq:HYP-def2}\\
\overline{V}_{n,\mu;\nu\,\rho} & = & \textrm{Proj}_{U(3)}[(1-\alpha_{3})
U_{n,\mu}+\frac{\alpha_{3}}{2}\sum_{\pm\eta\neq\rho,\nu,\mu}U_{n,\eta}U_{n+\hat{\eta},\mu}
U_{n+\hat{\mu},\eta}^{\dagger}]\,.\end{eqnarray}
is supplemented by the prescription (labeling lattice sites by $(x_0,\mathbf x)$)
\begin{eqnarray}
V_i(0,\mathbf x)\,=\widetilde V_i(0,\mathbf x)\,=\,\overline V_i(0,\mathbf x)&=&U_i(0,\mathbf x)\\
V_i(T,\mathbf x)=\widetilde V_i(T,\mathbf x)=\overline V_i(T,\mathbf x)&=&U_i(T,\mathbf x)
\end{eqnarray}
for the spatial links in the temporal boundaries. Here $U_i(0,\mathbf x)$ and $U_i(T,\mathbf x)$
are the prescribed spatial boundary fields and we use $\alpha_1=0.75,\,\alpha_2=0.6,\,\alpha_3=0.3$
as blocking parameters.
We note in passing that the classical background field induced by abelian homogeneous boundary
conditions is invariant under (iterated) smearing as long as a projection/normalization is
employed.

The improvement condition itself consists of demanding that $\Delta M$, the difference between two
definitions of a current quark mass, is equal to its tree--level value $\Delta M^{(0)}$ \cite{Luscher:1996ug}.
The cutoff effect $\Delta M$ is evaluated at the critical point (defined through yet another
current quark mass) and the clover coefficient is
tuned until this condition is satisfied.
Note that the use of a color background field in the SF allows quark mass definitions that
are $\rmO(a)$--improved without the knowledge of the axial current improvement coefficient $c_{\rm A}$
\cite{Luscher:1996ug}.
To summarize, at each value of the inverse bare gauge coupling $\beta$
a number of $c_{\rm SW}$ values are simulated, where in turn for each of those the hopping parameter
has to be tuned to find the critical point. An interpolation in $c_{\rm SW}$ then gives
the point where $\Delta M=\Delta M^{(0)}$.

The SF boundary improvement terms are dropped since they do not affect the observables
we will be interested in \cite{Luscher:1996sc}.
Lastly, physical units for the results are obtained using the hadronic scale $r_0$
\cite{Sommer:1993ce,Necco:2001xg} and assuming $r_0=0.5\,$fm.

\section{Numerical results}

All simulations are performed on lattices of size $8^3\times16$ and the background field
is chosen as in \cite{Luscher:1996ug}. First, we test the sensitivity of the improvement condition,
i.e. how $\Delta M$ varies with the clover coefficient, with results from $\beta=6.45$ shown as
filled symbols in \fig{fig:sens}. The point $c_{\rm SW}=1.043(11)$, where $\Delta M$ is equal to
its tree--level value $0.00028$, is also indicated. We obtain a clear signal for the clover
coefficient with a value that is rather close to one.

\DOUBLEFIGURE[b]{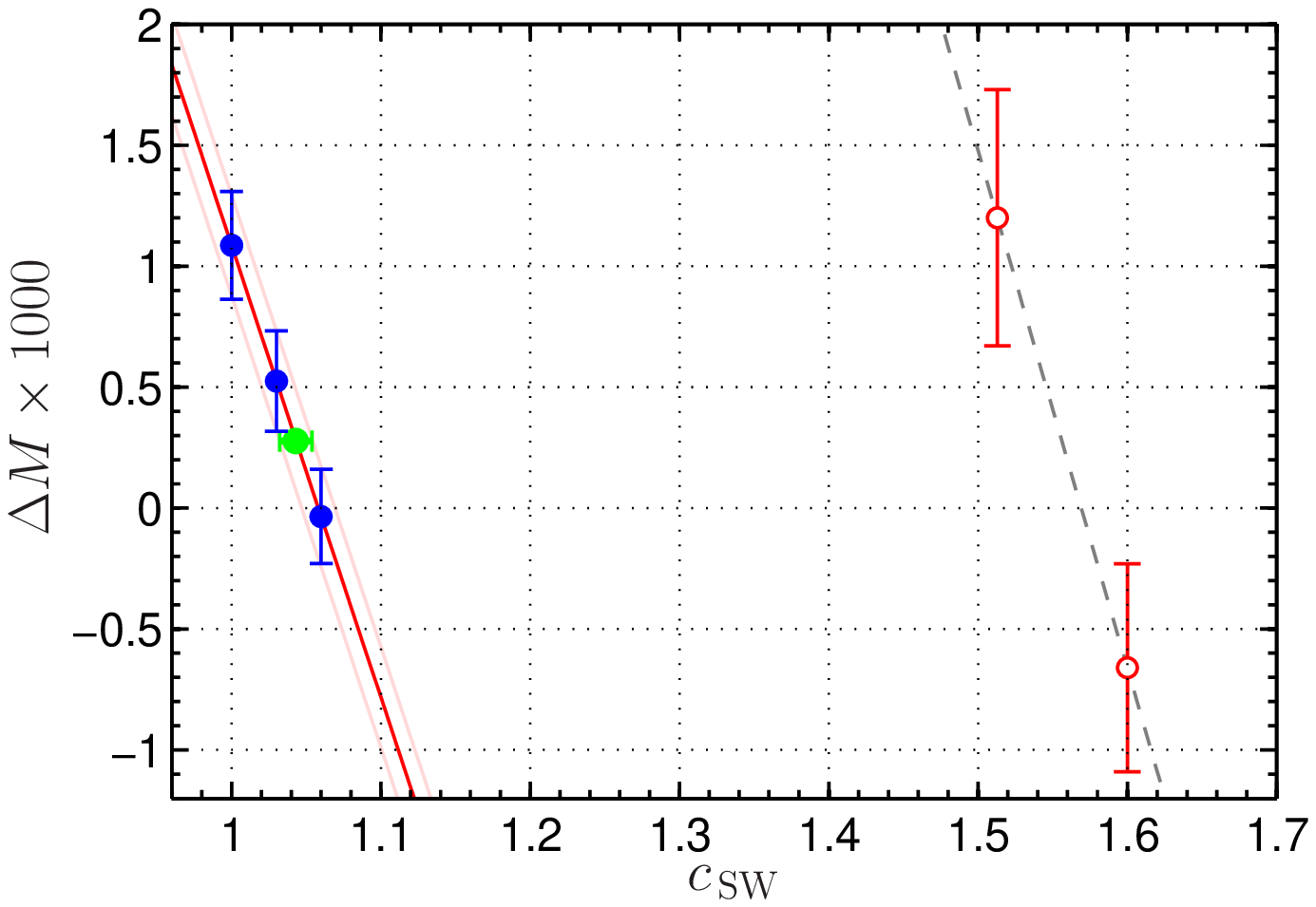,width=75mm}{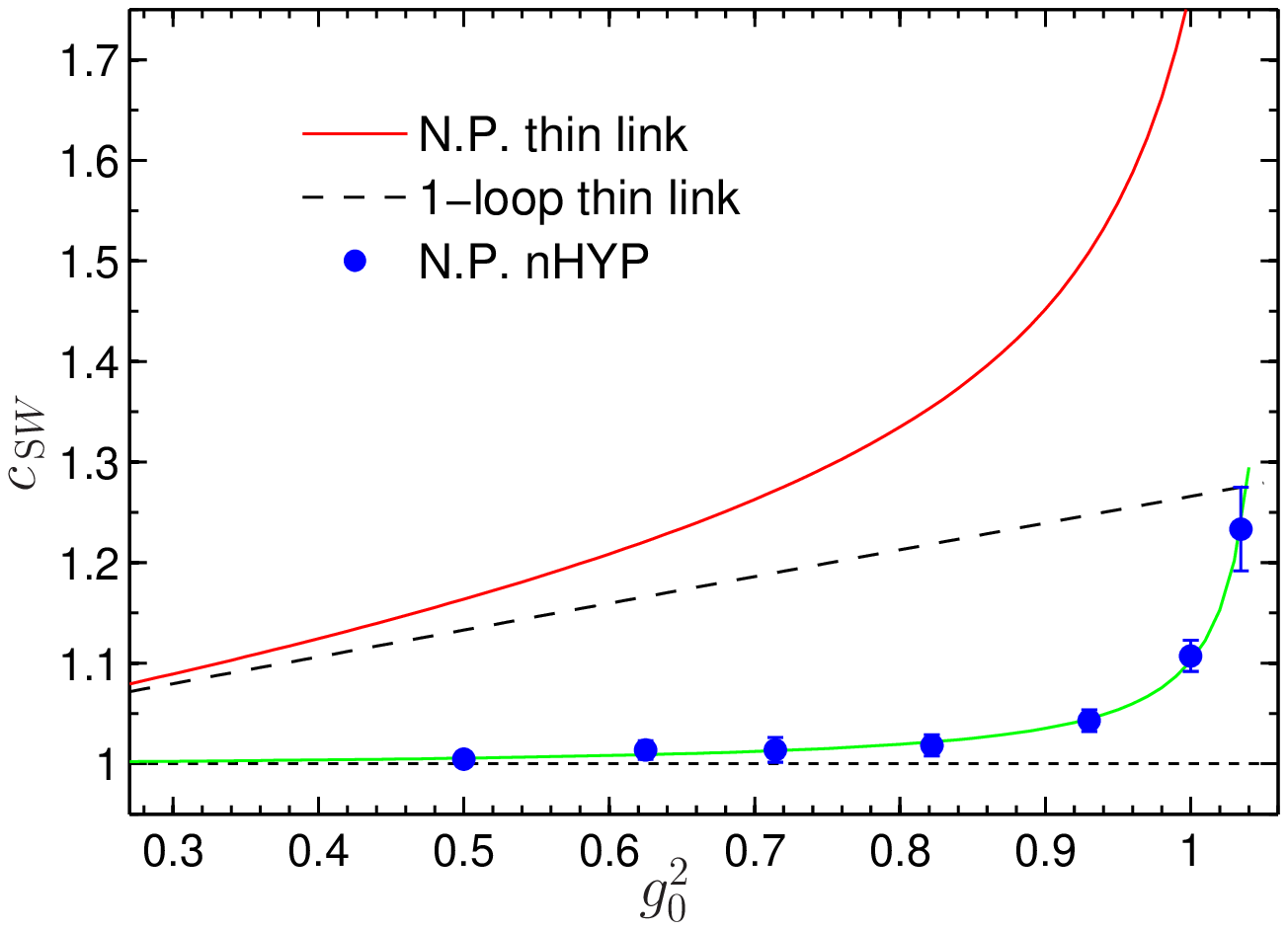,width=71.4mm}{
Sensitivity of the improvement condition: $\Delta M$ vs. $\csw$
for nHYP links at $\beta=6.45$ (filled symbols) and thin links at $\beta=6.4$ (open symbols,
\cite{Luscher:1996ug}).\label{fig:sens}}
{Non-perturbative result for $c_{\rm SW}$ with nHYP links and comparison to the thin
link results.\label{fig:csw}}

For comparison we also show results from \cite{Luscher:1996ug} at a similar lattice spacing
($\beta=6.4$, open symbols). Clearly, the sensitivity is very similar, while the
value of the clover coefficient, that is necessary to achieve non-perturbative improvement,
differs greatly.
In other words, the cutoff effects that are being canceled with a non-perturbative
$c_{\rm SW}$ are dramatically reduced if the fermions are coupled to the nHYP smeared gauge
field.

The determination was done for $5.8\leq\beta\leq12$ and the resulting non-perturbative
values for $c_{\rm SW}$ are plotted in \fig{fig:csw}. Note that, while the error
on our determination of the clover coefficient grows when one goes to coarser lattices
due to the increased statistical fluctuations,
the value itself rises only slowly. At $\beta=6$ we obtain $c_{\rm SW}=1.107(15)$ compared
to $c_{\rm SW}\simeq1.8$ with thin links \cite{Luscher:1996ug}.
Our data is well described by the interpolating formula
\be
c_{\,\mathrm SW}(\beta)=\frac{\beta-5.611}{\beta-5.647}\;.\label{eq1}
\ee

By design the result from the non-perturbative improvement condition will approach the
perturbative expansion for small enough value of $g_0^2$. From the plot it is then evident
that the 1--loop contribution to $c_{\rm SW}$ is very small in the case of nHYP smearing.
More precisely, from the asymptotic behavior of \eq{eq1} we would estimate
$c_{\rm SW}^{(1)}\simeq0.006$ as opposed to $c_{\rm SW}^{(1)}=0.266$ without smearing.

After fixing $c_{\rm SW}$ non-perturbatively Ref.~\cite{Luscher:1996ug} proceeds with
the improvement of the axial current, i.e. a determination of the coefficient $c_{\rm A}$.
Already with thin links the remaining cutoff effects are rather small and so is the
value of $c_{\rm A}$. Various methods to determine the axial current improvement coefficient
are available \cite{Durr:2003nc} and we have tested those that utilize variations in the
quark mass with respect to the insertion time and the periodicity angle $\theta$ of the
spatial boundary conditions.

No sensible criterion could be found since the observed cutoff effects were exceedingly
small, but a more detailed study and/or the use of wave functions
\cite{DellaMorte:2005se} might lead to better results. We therefore proceed
to test the quenched nHYP clover action with $c_{\rm SW}$ given by \eq{eq1}
and $c_{\rm A}=0$. The following simulations are performed with vanishing background
field.

\section{Testing the improved action}

\subsection{Remnant cutoff effects}

\DOUBLEFIGURE[b]{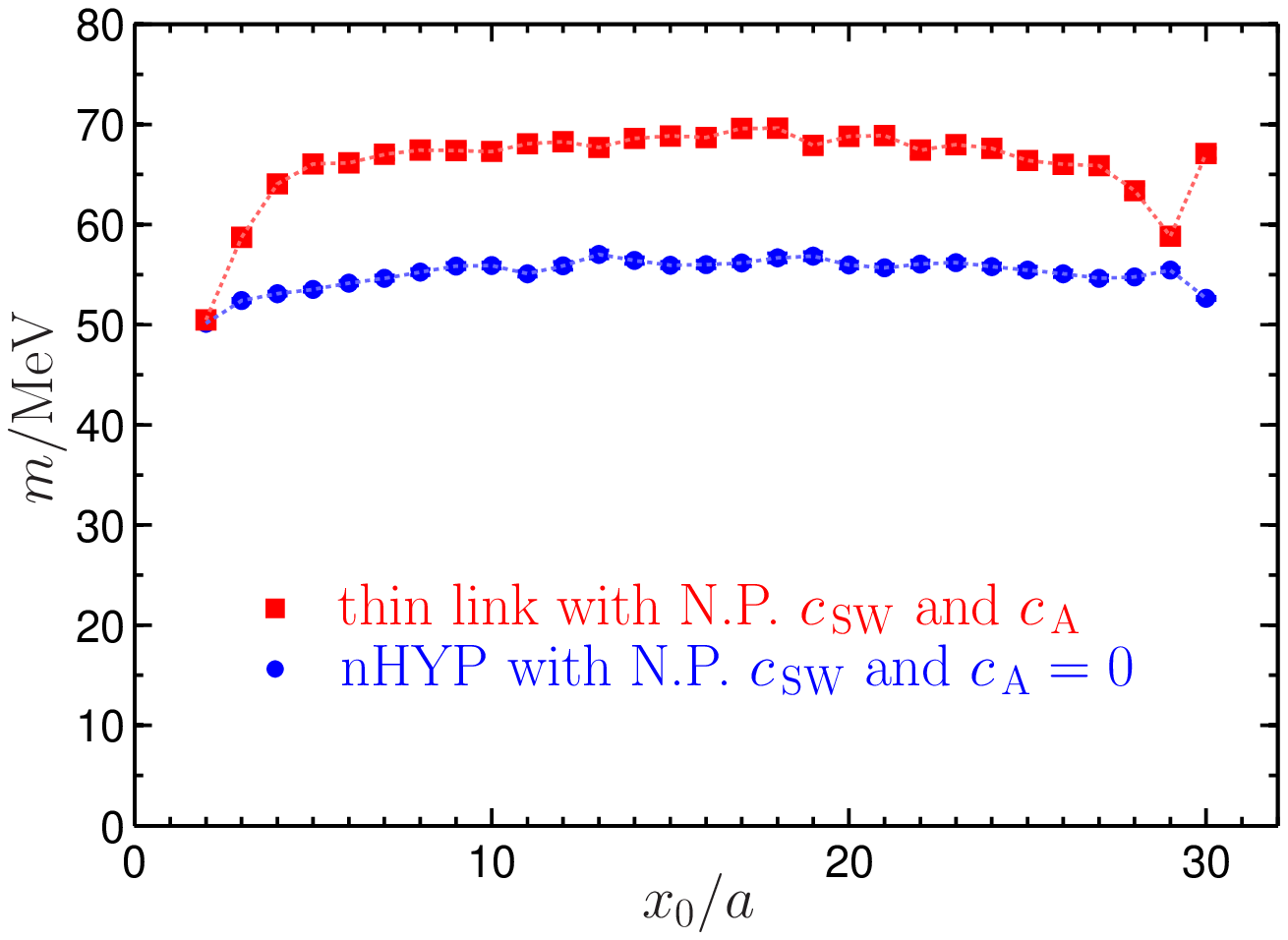,width=72mm}{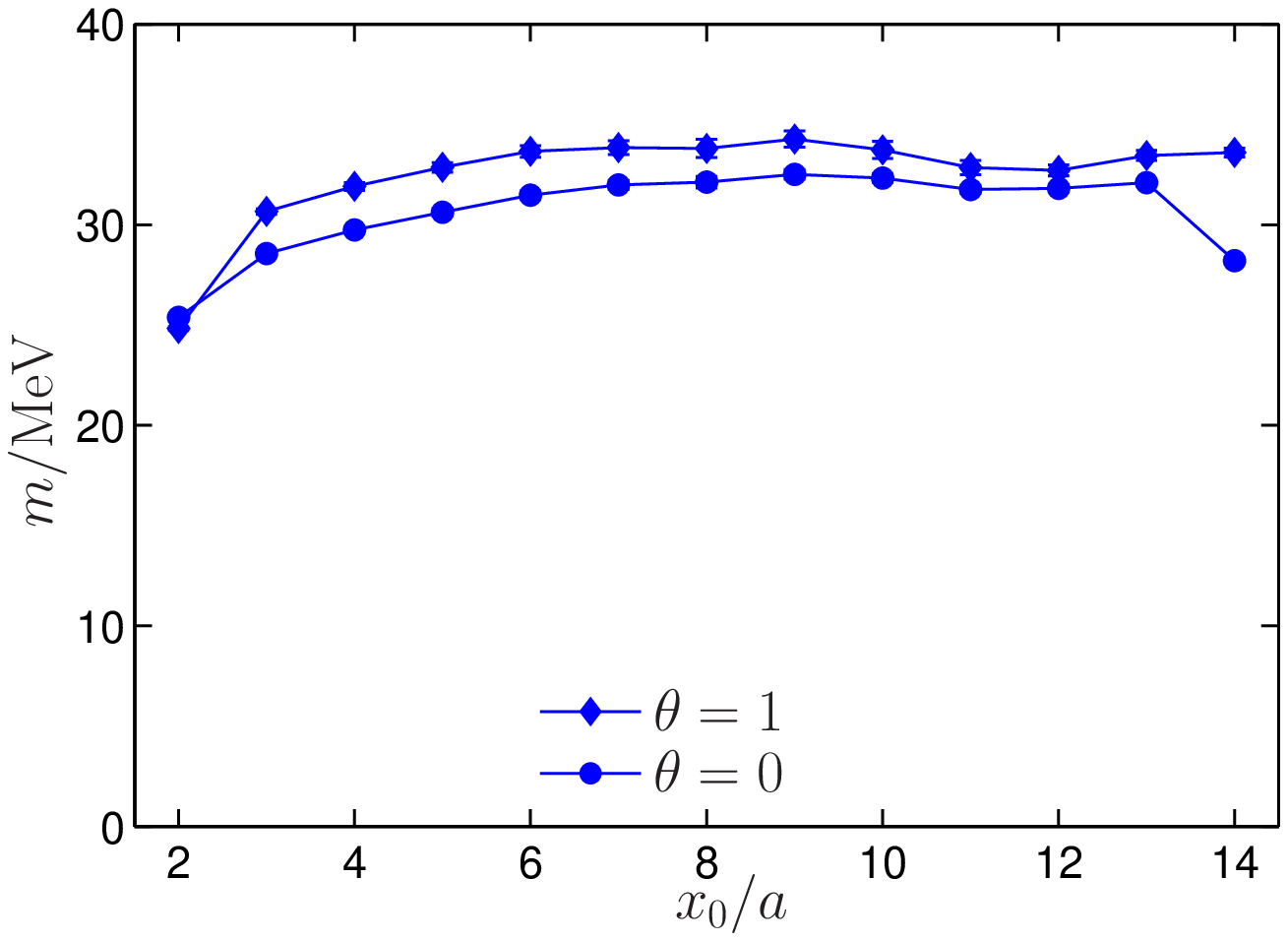,width=79.5mm}{
The PCAC mass at $\beta=6.2$ with and without nHYP smearing after non-perturbative
improvement.\label{fig:beta62}}
{The PCAC mass at $\beta\!=\!6$ with nHYP smearing
for two values of the spatial fermionic boundary phase.\label{fig:beta6}}

As a first test we consider the time dependence of the PCAC quark mass
on a $16^3\times32$ lattice at $\beta=6.2$, corresponding to a lattice spacing
of $a=0.068\,$fm. In \fig{fig:beta62} the result is compared to a
thin link simulation from Ref.~\cite{Luscher:1996ug} at a similar mass. In
both cases the quark mass is contained in a narrow band of $\simeq4$MeV width in the
interior of the lattice. However, toward the temporal boundaries the smeared action
clearly shows smaller cutoff effects since it remains much flatter than
in the thin link case.

Since for a smeared action this is an already rather fine lattice, we also perform
a test at $\beta=6$, $a\simeq0.1\,$fm, and $8^3\times16$.
\fig{fig:beta6} shows the PCAC quark mass with non-perturbative
$c_{\rm SW}$ at $\kappa=0.1257$ and $\theta=0$ and $1$.
By themselves, both data sets are contained in a $3\,$MeV band in the interior of
the lattice and even together, the spread is only $\simeq5\,$MeV.
For reference, note that the mass difference between $\theta=0$ and $1$ at $x_0=T/2$ in lattice
units is only 0.00079 or roughly twice its tree--level value.

\subsection{Vector current renormalization}

\EPSFIGURE[b]{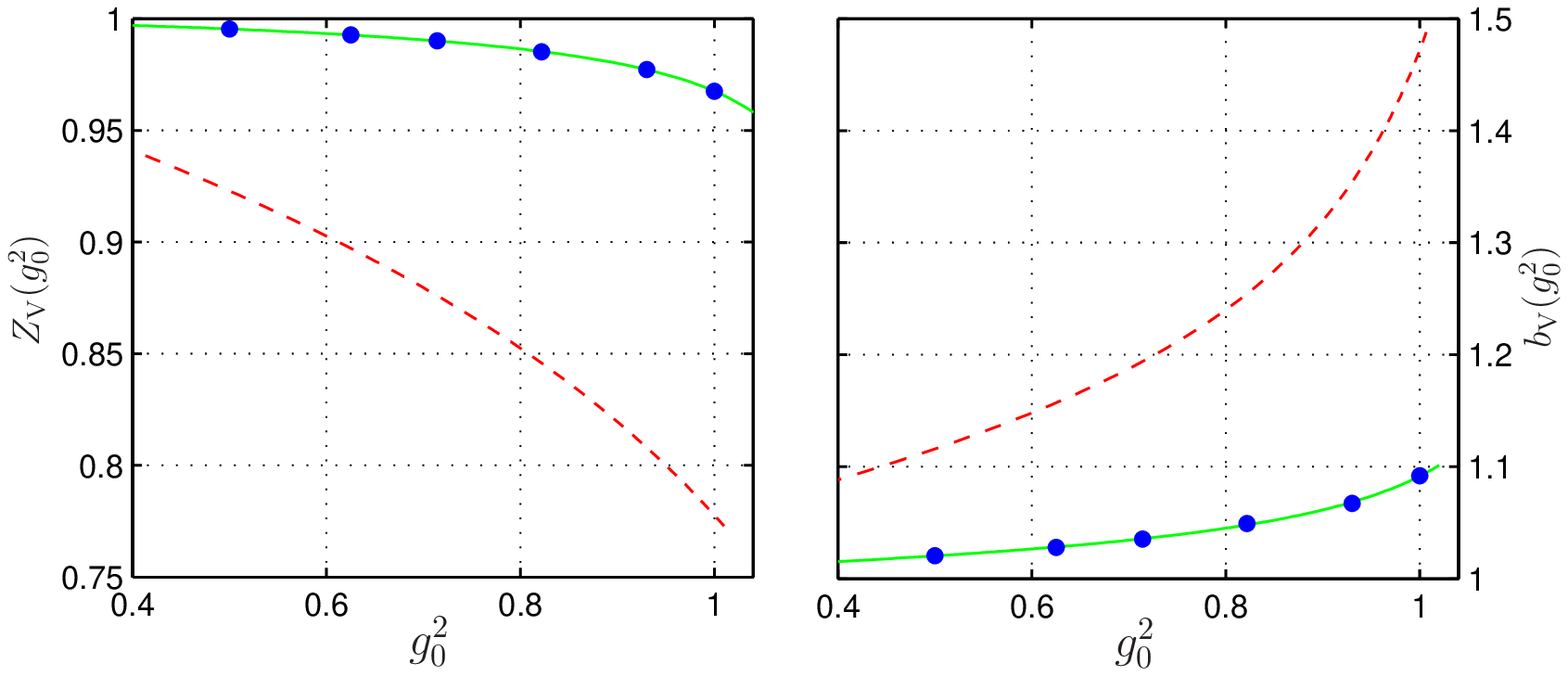,width=145mm}{
Non-perturbative results for $Z_{\rm V}$ and $b_{\rm V}$ for nHYP smeared Wilson fermions (filled circles,
error bars are smaller than symbol size).
For comparison the interpolating formulae for thin links from Ref.~\cite{Luscher:1996jn} are
shown as dashed lines.\label{fig:zv}}

Let us now consider the renormalization of the (local) isovector vector current.
In a mass--independent renormalization scheme the latter is given by
\be
(V_{\rm R})_\mu^a=Z_V(1+b_{\rm V}am_q)V_\mu^a\;,
\ee
where $V_\mu^a$ is the bare current, $m_q$ the bare subtracted quark mass and the term proportional
to $b_V$ is required to preserve
$\rmO(a)$--improvement under renormalization \cite{Luscher:1996sc}. Both $Z_{\rm V}$ and $b_{\rm V}$ have
a perturbative expansion of the form $1+\rmO(g_0^2)$. For the thin link case
the finite and scale--independent renormalization constant $Z_{\rm V}$ was calculated in \cite{Luscher:1996jn}
and found to deviate significantly from its tree--level value at relevant lattice spacings. The
same also holds for the dynamical case \cite{DellaMorte:2005rd}.

$Z_{\rm V}$ itself is obtained from a ratio of SF correlation functions at zero quark mass, while
$b_{\rm V}$ can be calculated from the slope of this ratio as a function of the subtracted
quark mass $m_q$ at $m_q\simeq0$. Results are shown as the filled symbols in \fig{fig:zv}.
Note that these were not obtained on a line of constant physics but at fixed $L/a$ and thus include a (presumably
small) ${\rmO}(a^2/L^2)$ uncertainty.
In both cases the results using nHYP smearing are significantly closer to unity than with thin
links. Even at a lattice spacing of $0.1\,$fm ($g_0^2=1$) we have $Z_V\simeq0.97$, indicating
that also the local vector current is almost conserved.

\subsection{Exceptional configurations}

Finally, we come to the issue of exceptional configurations. A tree--level analysis
\cite{Sint:1993un} has shown that the Schr\"odinger functional boundary conditions
induce an IR cutoff proportional to $1/T^2$ in the spectrum of the squared Hermitian
Dirac operator. But due to the absence of chiral symmetry, the Dirac operator
of the interacting theory can (and does) develop eigenvalues much smaller than this
bound. These lead to exceptional configurations where estimators for fermionic
correlation functions have values that are orders of magnitude above the normal
level of fluctuations.

In practice it was found \cite{Luscher:1996ug} that in the quenched SF with plaquette gauge
action $\beta\geq6.4$ is necessary to avoid exceptional configurations at zero quark mass.
Their occurrence is known to be linked to extremely localized
fluctuations of the gauge fields, so called dislocations. The Wilson operator, especially
with a clover term, is sensitive to those \cite{DeGrand:1998mn} and since the hypercubic
smearing was designed to optimally suppress dislocations, it should be well suited
to address this issue with Wilson clover fermions.

\EPSFIGURE[h]{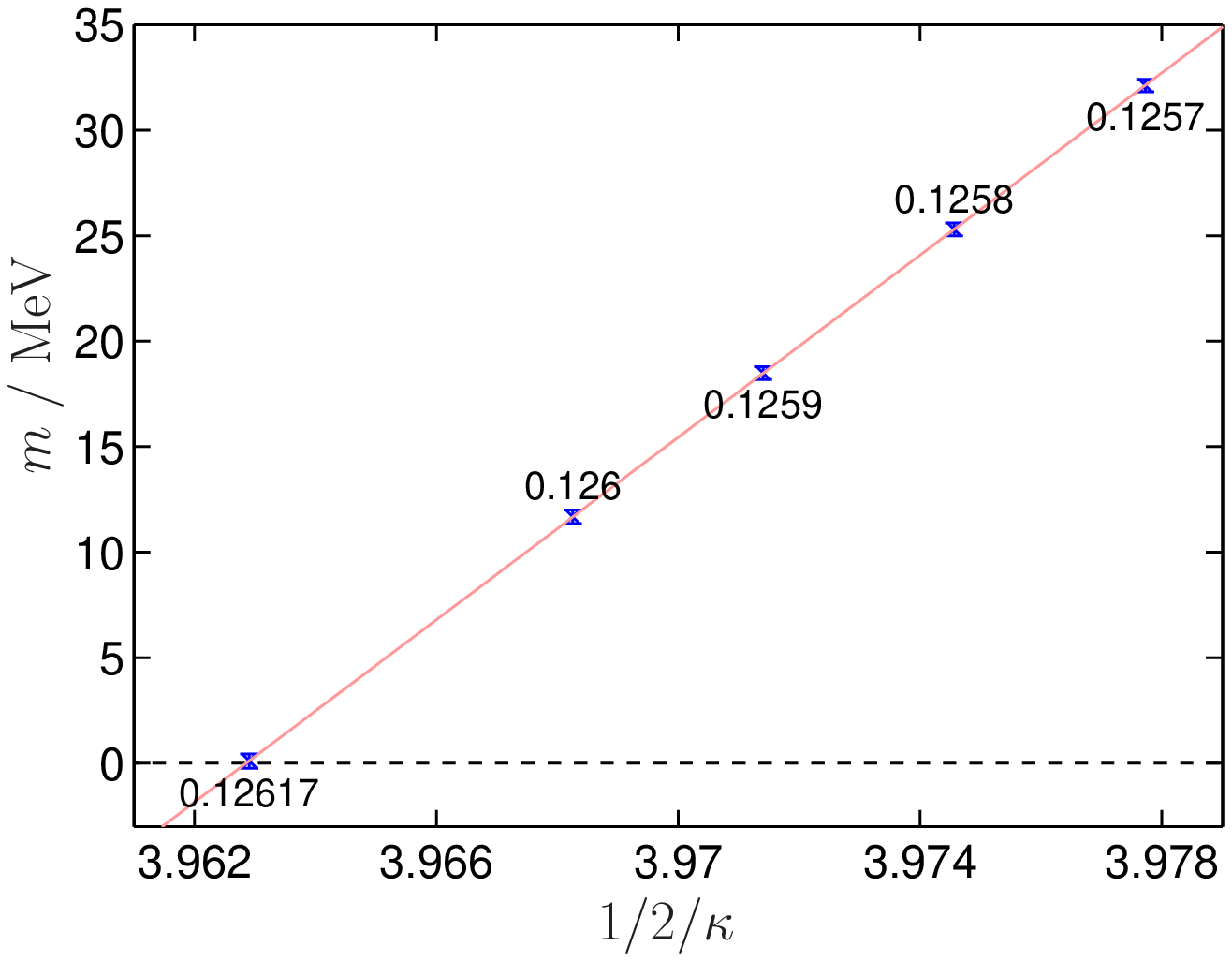,width=82mm}{
The current quark mass as a function of $(2\kappa)^{-1}$ with nHYP smearing at
$\beta=6$. The data points are labeled with the corresponding hopping parameter.\label{fig:crit}}

In \fig{fig:crit} we show the current quark mass as a function of $(2\kappa)^{-1}$ from
$8^3\times16$ simulations at $\beta=6$, $\theta=0$ and vanishing background field. No sign
of exceptional configurations was found in the statistical analysis. Repeating
this test on even coarser lattices ($\beta=5.8$, $a=0.136\,$fm) shows that there a quark mass
of about $25\,$MeV is required to avoid exceptional configurations. Strengthening
the SF cutoff with a smaller time extension $T/a\!=\!8$ again allows simulations
at the critical point even at this very coarse lattice. To summarize, we
estimate that in the SF the lattice spacing that is accessible along the critical line
is roughly doubled ($\beta=6$ vs. $6.4$) when nHYP smearing is employed.
We also note that at the same time the additive mass renormalization is
reduced by almost an order of magnitude.
At $\beta=6$ we obtain $\kappa_c=0.12617$, corresponding to $am_c=-0.037$,
whereas Ref.~\cite{Luscher:1996ug} quotes $\kappa_c=0.135196$, i.e. $am_c=-0.302$.

\section{Conclusion}

In this short study we have successfully used nHYP smeared clover fermions
in the framework of the Schr\"o\-dinger functional. We implemented the non-perturbative
action improvement pioneered by the ALPHA collaboration and found that no
large values of the clover coefficient are required to cancel the cutoff
effects under consideration. Instead, $c_{\rm SW}$ remains close to its
tree--level value even on coarse lattices.

As an example of a finite renormalization constant we performed the non-perturbative
renormalization of the local vector current. The renormalization constant $Z_{\rm V}$
differs from unity by only 3\% at a lattice spacing $a=0.1\,$fm.

We hope to have convinced the reader that also with modern (smeared) fermion formulations
the Schr\"odinger functional remains a useful tool for non-perturbative improvement and
renormalization.
Once a dynamical implementation is available, it will be very interesting to check to what
extent smearing helps with
respect to the large remnant $\rmO(a^2)$ effects and ambiguities found in two flavor clover simulations
\cite{DellaMorte:2005rd,Sommer:2003ne}.

This research was partially supported by the US Dept. of Energy.

{\renewcommand{\baselinestretch}{0.86}
\bibliography{../refs}
\bibliographystyle{JHEP-2}}

\end{document}